\documentclass[aps,showpacs,pra,notitlepage,twocolumn]{revtex4-2}
\usepackage{amssymb}
\usepackage{amsmath}
\usepackage{txfonts}
\usepackage{tikz,bm}
\usepackage{graphicx}
\usepackage{hyperref}
\usepackage{amsfonts}

\providecommand{\U}[1]{\protect\rule{.1in}{.1in}}
\renewcommand{\emph}[1]{\textcolor{blue}{\textit{#1}}}

\begin{document}
\title{Spin quantum number as quantum resource for quantum sensing}
\author{Qi Chai}
\affiliation{Beijing Computational Science Research Center, Beijing 100193, People's
Republic of China}
\author{Wen Yang}
\email{wenyang@csrc.ac.cn}
\affiliation{Beijing Computational Science Research Center, Beijing 100193, People's
Republic of China}

\begin{abstract}
Identifying quantum resources for quantum sensing is of paramount importance.
Up to date, two quantum resources has been widely recognized: the number $N$
of entangled quantum probes and the coherent evolution time $T$. Here we
identify the spin quantum number $S$ of high-spin systems as another quantum
resource, which can improve the sensing precision of magnetic field according
to the Heisenberg scaling in the absence of noises. Similar to the case of $N$
and $T$, the utility of $S$ may be degraded by environmental noises. We
analyze this point sysmatically under the Ornstein-Uhlenbeck noise (a
prevalent noise in realistic physical systems). We find that the utility of
$S$ vanishes in Markovian noises, but survives in non-Markovian noises, where
it improves the sensing precision according to the classical scaling
$1/\sqrt{S}$. Super-classical scaling can be achieved by suitable control of
the high-spin system.

\end{abstract}

\pacs{06.20.-f, 03.65.Ta, 07.55.Ge, 03.65.Yz}
% Metrology
% Foundations of quantum mechanics; measurement theory
% Magnetometers for magnetic field measurements
% Decoherence; open systems; quantum statistical methods

\maketitle

1. \textit{Introduction}. Quantum sensing as an important quantum technology
has attracted widespread interest in recent years, following the pioneering
works of Helstrom
\cite{helstrom1967minimum,helstrom1968minimum,helstrom1969quantum} and Holevo
\cite{holevo1973statistical}. The key idea is to utilize quantum resources
(such as $N$-particle entanglement) to improve the sensing precision beyond
its classical counterpart, where $N$ repeated measurements improves the
sensing precision by a factor $\sqrt{N}$ according to the central limit
theorem. In the absence of noises, proper use of $N$-particle entanglement
allows the sensing precision to attain the Heisenberg scaling $1/N$
\cite{giovannetti2006quantum,PhysRevA.94.012339}. The presence of
environmental noises limits the utility of $N$-particle entanglement and hence
degrades the sensing precision significantly: (1) In the worse case of
independent Markovian noise on each quantum probe \cite{huelga1997improvement}%
, the scaling of the sensing precision with respect to $N$ reduces to the
classical limit $1/\sqrt{N}$. (2) In the case of non-Markovian noise, the
sensing precision can maintain super-classical scaling $1/N^{3/4}$ with
respect to $N$ by suitable design of the evolution time
\cite{PhysRevA.94.012339,PhysRevLett.109.233601}. Recent breakthrough on the
theoretical side includes the use of suitable dynamical control, quantum
feedback, and quantum error correction to restore the Heisenberg scaling in
Markovian and non-Markovian noises \cite{Wang_2017, PhysRevA.102.060201, PhysRevLett.131.050801}. On the experimetal side, the advantages of
quantum sensing have been demonstrated in a variety of platforms, including
optical systems \cite{xiang2011entanglement, okamoto2008beating, chen2018heisenberg, PhysRevLett.126.070503, PhysRevLett.123.040501}, atomic
systems \cite{taylor2008high,chalopin2018quantum}, superconducting circuits
\cite{wang2019heisenberg}, and others.

Since quantum resources play a central role in quantum sensing, identifying
possible quantum resources is crucial. Up to date, two quantum resources have
been widely recognized: the number $N$ of entangled quantum probes and the
coherent evolution time $T$. Here we identify the spin quantum number $S$ of
high-spin systems as a new quantum resource for the accurate measurement of
magnetic field, which forms the cornerstone of many disciplines (magnetic
resonance imaging, magnetic navigation, material analysis, etc.)\ and has
become a prominent research focus \cite{henrichsen1998overview}. In the
absence of noises, the proper use of $S$ allows Heisenberg scaling $1/S$ for
the sensing precision, similar to the case of $N$ and $T$. In environmental
noises, the utility of $S$ still survives, and super-classical scaling
$1/S^{\alpha}$ ($\alpha>1/2$) can be maintained by suitable dynamical control
of the high-spin system. Since high-spin systems are ubiquitous, our work
enhances the power of quantum sensing.

2. \textit{Procedures for quantum sensing }The typical procedures for
estimating an unknown real parameter $\lambda$ using a quantum probe consists
of three steps \cite{zhang2018improving}:

(1) The quantum probe is prepared into certain initial state $\rho$ and then
undergoes certain $\lambda$-dependent evolution into a final state
$\rho_{\lambda}$ parametrized by the parameter $\lambda$ to be estimated;

(2) The quantum probe undergoes a general measurement described by a set of
positive operators $\{\Pi_{x}\}$ satisfying $\sum_{x}\Pi_{x}=1$. The
measurement outcome $x$ is a random variable sampled from the probability
distribution $P_{\lambda}\left(  x\right)  =\mathrm{Tr}\Pi_{x}\rho_{\lambda}$
that depends on $\lambda$ \cite{toth2014quantum}. Repeating the
initialization-evolution-measurement cycle $\nu$ times yields $\nu$
independent measurement data $\mathbf{x}\equiv(x_{1},\cdots,x_{\nu})$.

(3) Given $\mathbf{x}=(x_{1},\cdots,x_{\nu})$, an unbiased estimator
$\hat{\lambda}\left(  \mathbf{x}\right)  $ to $\lambda$ is constructed. Here
$\hat{\lambda}\left(  \mathbf{x}\right)  $ is a real-valued function of $x$,
hence is also a random variable. With $\mathbf{E}[\cdots]\equiv\sum_{x}%
(\cdots)P_{\lambda}(x)$, the unbiased condition reads $\lambda=\mathbf{E[}%
\hat{\lambda}\left(  x\right)  ]$ and the sensing precision/error $\delta
\hat{\lambda}\equiv\sqrt{\mathrm{var}(\hat{\lambda})}$ is defined as the
square root of the variance $\mathrm{var}(\hat{\lambda})$ of $\hat{\lambda}$.

The estimation precision of any unbiased estimator $\hat{\lambda}(\mathbf{x})$
based on $\nu$ repeated measurement outcomes $\mathbf{x}\equiv(x_{1}%
,\cdots,x_{\nu})$ sampled from $P_{\lambda}\left(  x\right)  $ obeys the
Cram\'{e}r-Rao bound $\delta\hat{\lambda}\geq1/\sqrt{\nu F}$
\cite{cramer1999mathematical}, where $F\equiv\mathbf{E}[L_{\lambda}^{2}(x)]$
is the (classical) Fisher information (CFI) from one datum $x$ and
$L_{\lambda}(x)\equiv\partial\ln P_{\lambda}\left(  x\right)  /\partial
\lambda$ is the (classical) score of the distribution $P_{\lambda}(x)$. The
quantum analog to the classical score $L_{\lambda}(x)$ is the quantum score
$L_{\lambda}$, an Hermitian operator satisfying $\partial_{\lambda}%
\rho_{\lambda}=\left(  L_{\lambda}\rho_{\lambda}+\rho_{\lambda}L_{\lambda
}\right)  /2$ \cite{paris2009quantum}. The CFI $F$ associated with the
distribution $P_{\lambda}(x)$ of any measurement $\{\Pi_{x}\}$ performed on
$\rho_{\lambda}$ is upper bounded by the quantum Fisher information (QFI)
$\mathcal{F}\equiv\mathrm{Tr}\rho_{\lambda}L_{\lambda}^{2}$
\cite{paris2009quantum}. This leads to the quantum Cram\'{e}r-Rao bound:
\[
\delta\hat{\lambda}\geq\frac{1}{\sqrt{\nu F}}\geq\frac{1}{\sqrt{\nu
\mathcal{F}}}.
\]
Here we focus on local estimation problems, where a sufficiently accurate
estimator $\lambda_{0}$ to $\lambda$ is known prior to the measurement. In
this case, both inequalities above can be saturated simultaneously, i.e.,
there exists an optimal measurement making $F=\mathcal{F}$ (such as the
projective measurement over the quantum score $L_{\lambda_{0}}$, see Ref.
\cite{yang2019optimal} and references therein) and there exists an
\textit{efficient} unbiased estimator $\hat{\lambda}$ with error $\delta
\hat{\lambda}=1/\sqrt{\nu F}=1/\sqrt{\nu\mathcal{F}}$. Therefore, the QFI
determines the ultimate achievable estimation precision $\delta\hat{\lambda
}_{\min}=1/\sqrt{\nu\mathcal{F}}$.

Next we show that the spin quantum number $S$ $(S=1/2,$ $1,$ $3/2,$ $\cdots$)
of a high-spin system can be utilized as a quantum resource to improve the
sensing precision of a magnetic field $B$ (i.e., the parameter to be estimated
is $\lambda=B$).

3. \textit{Noise-free magnetometry. }Under the static magnetic field $B$ to be
measured, the Hamiltonian of the spin-$S$ probe is $H=\gamma BS_{z}%
\equiv\omega S_{z}$, where $\gamma$ is the gyromagnetic ratio. For brevity, we
set $\gamma=1$, so that the parameter to be estimated is the Larmor frequency
$\omega=B$. Starting from $\rho(0)$, the evolution for an interval $\tau$
causes Larmor precession of the spin about the $z$ axis by an angle
$\omega\tau=\gamma B\tau$ and the final state is $\rho_{\omega}(\tau
)=e^{-i\omega\tau S_{z}}\rho(0)e^{i\omega\tau S_{z}}$. For a fixed $\tau$, the
optimal initial state \cite{giovannetti2006quantum, yang2019optimal}
\begin{equation}
\left\vert \psi\right\rangle =\frac{1}{\sqrt{2}}\left(  \left\vert
S\right\rangle +\left\vert -S\right\rangle \right)  \label{GHZ}%
\end{equation}
leads to maximal QFI
\begin{equation}
\mathcal{F}=(2S)^{2}\tau^{2}\label{QFI_NOISEFREE}%
\end{equation}
in the final state $|\psi_{\omega}(\tau)\rangle\equiv e^{-i\omega\tau S_{z}%
}|\psi\rangle$, where $|\pm S\rangle$ are eigenstates of $S_{z}$ with
eigenvalues $\pm S$. The ultimate\ sensing precision after $\nu$ repeated
measurements is
\[
\delta\omega_{\min}=\frac{1}{\sqrt{\nu\mathcal{F}(\tau)}}=\frac{1}{\sqrt{\nu}%
}\frac{1}{(2S)\tau}.
\]
Therefore, the spin quantum number $S$ is a quantum resource capable of
improving the sensing precision according to the Heisenberg scaling $1/S$, in
analogy to the number $N$ of entangled quantum probes \cite{giovannetti2006quantum,PhysRevA.94.012339}.

4. \textit{Spin decoherence under magnetic noise. }In the presence of a
magnetic noise $\tilde{\omega}(t)$ (with zero mean) along the $z$ axis, the
total magnetic field along the $z$ axis becomes $\omega+\tilde{\omega}(t)$
(recall we set $\gamma=1$ for conciseness) and the Hamiltonian of the spin-$S$
becomes $H(t)=\left[  \omega+\tilde{\omega}\left(  t\right)  \right]  S_{z}$
\cite{zhang2018improving}. We take the GHZ-like state Eq. (\ref{GHZ}) as the
initial state. This allows us to obtain explicit analytical expressions and
hence a clear physical picture about the utility of $S$ in different regimes
of the noise. The optimal initial state in the noisy case will be discussed shortly.

After an interval $\tau$, the final state is
\[
|\psi_{\omega}(\tau)\rangle=\frac{1}{\sqrt{2}}\left(  e^{-iS\omega\tau
}e^{-iS\tilde{\varphi}(\tau)}\left\vert S\right\rangle +e^{iS\omega\tau
}e^{iS\tilde{\varphi}(\tau)}\left\vert -S\right\rangle \right)  ,
\]
where $\tilde{\varphi}(\tau)\equiv\int_{0}^{\tau}\tilde{\omega}(t)dt$ is the
accumulated random phase due to the noise. The density matrix of the spin-$S$
is%
\begin{multline*}
\rho_{\omega}\left(  \tau\right)  =\overline{|\psi_{\omega}(\tau
)\rangle\langle\psi_{\omega}(\tau)|}=\frac{1}{2}\left(  \left\vert
S\right\rangle \left\langle S\right\vert +\left\vert -S\right\rangle
\left\langle -S\right\vert \right)  \\
+\frac{1}{2}\left(  e^{-i(2S)\omega\tau}\overline{e^{-i(2S)\tilde{\varphi
}(\tau)}}\left\vert S\right\rangle \left\langle -S\right\vert +h.c.\right)  ,
\end{multline*}
where $\overline{(\cdots)}$ stands for the average over the distribution of
the noise $\tilde{\omega}(t)$. To proceed, we consider a widely encountered
noise in realistic physical systems: the Ornstein--Uhlenbeck noise generated
by the Ornstein--Uhlenbeck stochastic process \cite{benedetti2014characterization}. Since the Ornstein--Uhlenbeck
noise is Gaussian, it is completely specified by its auto-correlation function
$C\left(  t,t^{\prime}\right)  =b^{2}e^{-\left\vert t-t^{\prime}\right\vert
/\tau_{c}}$, which contains two parameters: $b$ is the magnitude of the noise
and $\tau_{c}$ is the memory time of the noise. Since $\tilde{\omega}(t)$ is
Gaussian, the random phase $\tilde{\varphi}$ is also Gaussian. Using
$\overline{e^{-iX}}=e^{-i\overline{X}}e^{-\mathrm{var}(X)/2}$ for any Gaussian
random variable $X$, we have
\begin{eqnarray*}
\rho_{\omega}\left(  \tau\right)&=&\frac{1}{2}\left(  \left\vert
S\right\rangle \left\langle S\right\vert +\left\vert -S\right\rangle
\left\langle -S\right\vert \right)  \\
&& +\frac{1}{2}\left(  e^{-i(2S)\omega\tau}e^{-(2S)^{2}\chi(\tau)}\left\vert
S\right\rangle \left\langle -S\right\vert +h.c.\right)  ,
\end{eqnarray*}
where
\[
\chi(\tau)\equiv\frac{1}{2}\overline{\tilde{\varphi}^{2}(\tau)}=b^{2}\tau
_{c}^{2}\left(  \frac{\tau}{\tau_{c}}+e^{-\tau/\tau_{c}}-1\right)
\]
accounts for the noise-induced decoherence of the spin-$S$, i.e., decay of the
off-diagonal coherence $\left\vert \langle S|\rho_{\omega}\left(  \tau\right)
|-S\rangle\right\vert =e^{-\left(  2S\right)  ^{2}\chi(\tau)}$.

When $\tau$ increases, $\chi(\tau)$ increases monotonically:
\begin{equation}
\chi\left(  \tau\right)  \approx\left\{
\begin{array}
[c]{ll}%
\frac{1}{2}b^{2}\tau^{2} & \text{ \ }\left(  \tau\ll\tau_{c}\right)  ,\\
b^{2}\tau_{c}\tau & \text{ \ }\left(  \tau\gg\tau_{c}\right)  .
\end{array}
\right.  \label{KAI_LIMIT}
\end{equation}
As a function of $\tau$, the off-diagonal coherence $\left\vert \langle
S|\rho_{\omega}\left(  \tau\right)  |-S\rangle\right\vert =e^{-\left(
2S\right)  ^{2}\chi(\tau)}$ goes from Gaussian decay in the short-time regime
$\tau\ll\tau_{c}$ to the exponential decay in the long-time regime $\tau
\gg\tau_{c}$. The spin decoherence time $T_{2}$ is usually $T_{2}$ defined as
the time at which the off-diagonal coherence decays to $1/e$ of its initial
value:
\begin{equation}
\left(  2S\right)  ^{2}\chi(T_{2})=1.
\label{spin decoherence time}
\end{equation}
When $\tau_{c}\ll T_{2}$, we say the noise is Markovian, because the Gaussian
decoherence extending from $\tau=0$ to $\tau\sim\tau_{c}$ is negligibly narrow
compared with the exponential decoherence extending from $\tau\sim\tau_{c}$ to
$\tau\gtrsim T_{2}$; otherwise we say the noise is non-Markovian. In the
extreme case $T_{2}\ll\tau_{c}$, we say the noise is quasi-static. Using Eq.
(\ref{KAI_LIMIT}), we obtain%
\begin{equation}
T_{2}\approx\left\{
\begin{array}
[c]{ll}%
\frac{1}{\sqrt{2}Sb}\ll\tau_{c} & \ (\text{Quasi-static: }2Sb\tau_{c}\gg1),\\
\frac{1}{(2Sb)^{2}\tau_{c}}\gg\tau_{c} & \ \left(  \text{Markovian: }%
2Sb\tau_{c}\ll1\right)  .
\end{array}
\right.  \label{T2}%
\end{equation}
Why $2Sb\tau_{c}$ determines the Markovianity of the noise? Intuitively, a
noise $\tilde{\omega}(t)$ with magnitude $b$ and memory time $\tau_{c}$ can be
roughly viewed as a random variable jumping between $\pm b$ at an interval
$\tau_{c}$. Therefore, $2Sb\tau_{c}$ is the noise-induced relative phase
between $|\pm S\rangle$ in one jumping interval $\tau_{c}$, while $T_{2}$ is
roughly the time cost for this relative phase to attain unity:

(1) If the relative phase accumulated in one jumping interval is small
$2Sb\tau_{c}\ll1$, then many jumping intervals are necessary for this relative
phase to accumulate to unity, i.e., $T_{2}\gg\tau_{c}$.

(2) If the relative phase accumulated in one jumping interval is large
$2Sb\tau_{c}\gg1$, then only a small fraction of one jumping interval
$\tau_{c}$ is necessary for this relative phase to accumulated to unity, i.e.,
$T_{2}\ll\tau_{c}$.

Therefore, whether the noise is Markovian or not is determined by not only the
noise parameters $b,\tau_{c}$ themselves, but also the magnitude of the spin
quantum number $S$.\ Interestingly, since the whole evolution of the spin-$S$
only involves two basis states $\left\vert S\right\rangle $ and $\left\vert
-S\right\rangle $, we can establish a connection between the spin-$S$
magnetometry and spin-1/2 magnetometry by identifying $\left\vert
S\right\rangle $ ($\left\vert -S\right\rangle $) as the spin-up (spin-down)
state of the spin-1/2: $\left\vert \uparrow\right\rangle \equiv\left\vert
S\right\rangle $ and $\left\vert \downarrow\right\rangle \equiv\left\vert
-S\right\rangle $ and identifying $s_{z}\equiv S_{z}/(2S)$ as a spin-1/2
operator. Consequently, the Zeeman Hamiltonian of the spin-$S$ can be written
as
\[
H(t)=\left[  \omega+\tilde{\omega}\left(  t\right)  \right]  S_{z}=2S\left[
\omega+\tilde{\omega}\left(  t\right)  \right]  s_{z}.
\]
In other words, using a spin-$S$ instead of a spin-1/2 for magnetometry
amounts to enhancing the signal $\omega$ and the noise $\tilde{\omega}\left(
t\right)  $ (hence its magnitude $b$) by the same factor $2S$.

\begin{figure}[ptb]
\centering
\includegraphics[width=\columnwidth]{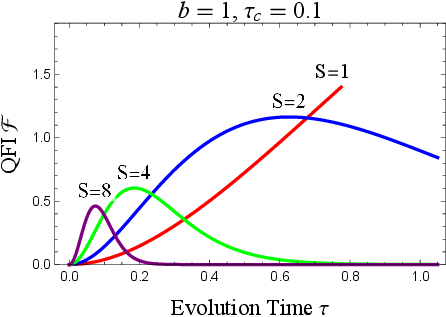}
\caption{The variation of QFI of spin-S system with evolution time $\tau$ when $b=1, \tau_{c}=0.1$}%
\label{fig:qfi with tau 1}%
\end{figure}

5. \textit{Noisy magnetometry. }The QFI of the final state $\rho_{\omega
}\left(  \tau\right)  $ is
\begin{equation}
\mathcal{F}(\tau)=\left(  2S\right)  ^{2}\tau^{2}e^{-2\left(  2S\right)
^{2}\chi(\tau)}\label{QFI}%
\end{equation}
and the ultimate sensing precision is%
\begin{equation}
\delta\omega_{\min}=\frac{1}{\sqrt{\nu\mathcal{F}(\tau)}}=\frac{1}{\sqrt{\nu}%
}\frac{1}{(2S)\tau e^{-\left(  2S\right)  ^{2}\chi(\tau)}},\label{DWMIN}%
\end{equation}
where the notation $\mathcal{F}(\tau)$ emphasizes that the final-state QFI
depends on the evolution time $\tau$. Compared with the noise-free case, the
noise gives rise to the decoherence factor $\left\vert \langle S|\rho_{\omega
}\left(  \tau\right)  |-S\rangle\right\vert =e^{-\left(  2S\right)  ^{2}%
\chi(\tau)}$ that degrades the the sensing precision. Due to the competition
between the signal accumulation factor $\tau$ and the spin decoherence factor
$e^{-\left(  2S\right)  ^{2}\chi(\tau)}$, the estimation error in Eq.
(\ref{DWMIN}) as a function of $\tau$ shows a minimum at $\tau\sim T_{2}$ (cf.
Fig. \ref{fig:qfi with tau 1}):%
\[
\tau\approx\left\{
\begin{array}
[c]{ll}%
\dfrac{1}{2Sb}=\dfrac{T_{2}}{\sqrt{2}}\ll\tau_{c} & \ (\text{Quasi-static:
}2Sb\tau_{c}\gg1)\\
\dfrac{1}{(2Sb)^{2}\tau_{c}}=T_{2}\gg\tau_{c} & \ \left(  \text{Markovian:
}2Sb\tau_{c}\ll1\right)
\end{array}
\right.  .
\]
Intuitively, when $\tau\ll T_{2}$, the spin decoherence factor $e^{-\left(
2S\right)  ^{2}\chi(\tau)}\approx1$ and the signal accumulation factor
dominates, so increasing $\tau$ improves the estimation precision. When
$\tau\gg T_{2}$, the rapid decay of the spin decoherence factor $e^{-\left(
2S\right)  ^{2}\chi(\tau)}$ dominates, so increasing $\tau$ degrades the
estimation precision. Therefore, best estimation precision always occurs at
$\tau\sim T_{2}$.

In Fig. \ref{fig:qfi with tau 1}, the QFI for $S=4$ increases to its maximum
value $\mathcal{F}\approx0.6$ after an evolution time $\tau\approx0.2$; while
the QFI for $S=8$ increases to its maximum value $\mathcal{F}\approx0.45$
after an evolution time $\tau\approx0.07$. Which case has a better
performance? Although the case $S=4$ gives a maximum QFI larger than the case
$S=8$, it also consumes more evolution time, which is also a valuable quantum resource.

To make a fair comparison, we consider the following task: given a spin-$S$
quantum probe and the total time cost $T$, find the optimal estimation
strategy that gives the highest precision, i.e., the smallest estimation
error. When $T/T_{2}$ is large, the optimal strategy is to equally divide the
total time cost into $\nu$ intervals of duration $\tau=T/\nu$ (the value of
$\tau$ is to be optimized) and perform one measurement in each interval. After
$\nu$ repeated measurements, the estimation precision follows from the quantum
Cram\'{e}r-Rao bound as%
\[
\delta\omega_{\min}=\frac{1}{\sqrt{\nu\mathcal{F}\left(  \tau\right)  }}%
=\frac{1}{\sqrt{T}}\frac{1}{\sqrt{\mathcal{F}\left(  \tau\right)  /\tau}},
\]
where $\mathcal{F}\left(  \tau\right)  $ is the QFI after an evolution time
$\tau$ [see Eq. (\ref{QFI})]. Obviously, the estimation precision improves
when the total time cost $T$ increases. Therefore, instead of working with the
estimation precision, it is more convenient to work with the estimation
precision per unit time:%
\[
\delta\omega_{\min}\sqrt{T}=\frac{1}{\sqrt{\mathcal{F}\left(  \tau\right)
/\tau}},
\]
where $\mathcal{F}\left(  \tau\right)  /\tau$ is the QFI yield rate. Next we
need only tune $\tau$ to maximize the QFI yield rate. Generally, this can only
be done by numerical optimization. However, in the extreme case of
quasi-static noise and Markovian noise, we can obtain analytical expressions
for the (optimized) QFI yield rate,%
\[
R\equiv\max_{\tau}\frac{\mathcal{F}\left(  \tau\right)  }{\tau}=\left\{
\begin{array}
[c]{ll}%
\sqrt{\dfrac{2}{e}}\dfrac{S}{b}=\dfrac{2}{\sqrt{e}}S^{2}T_{2} & \ \left(
\text{Quasi-static noise}\right)  \\
\dfrac{1}{2e}\dfrac{1}{b^{2}\tau_{c}}=\dfrac{2}{e}S^{2}T_{2} & \ \left(
\text{Markovian noise}\right)
\end{array}
\right.  ,
\]
and the optimal evolution time%
\begin{equation}
\tau_{\mathrm{opt}}=\left\{
\begin{array}
[c]{cc}%
\frac{1}{\sqrt{2}2Sb} & \ \left(  \text{Quasi-static noise}\right)  \\
\frac{1}{2\left(  2Sb\right)  ^{2}\tau_{c}} & \ \left(  \text{Markovian
noise}\right)
\end{array}
\right.  =\frac{T_{2}}{2}.\label{tau_opt}%
\end{equation}
Since $\tau_{\mathrm{opt}}\sim T_{2}$, the QFI yield rate
\begin{equation}
R\sim\frac{\mathcal{F}\left(  T_{2}\right)  }{T_{2}}\sim\left(  2S\right)
^{2}T_{2}\label{optimized QFI yield rate}%
\end{equation}
can be obtained from $\mathcal{F}\left(  \tau\right)  /\tau=\left(  2S\right)
^{2}\tau$ for the noise-free case by replacing $\tau\rightarrow T_{2}$. This
is because in the noisy case, once $\tau$ exceeds $\sim T_{2}$, the rapid
decay of the decoherence factor dominates and begins to degrade the estimation
precision when $\tau$ further increases. In other words, the existence of the
noise amounts to setting an upper bound $T_{2}$ on the coherent evolution time
$\tau$.

\begin{figure}[ptb]
\centering
\includegraphics[width=\columnwidth]{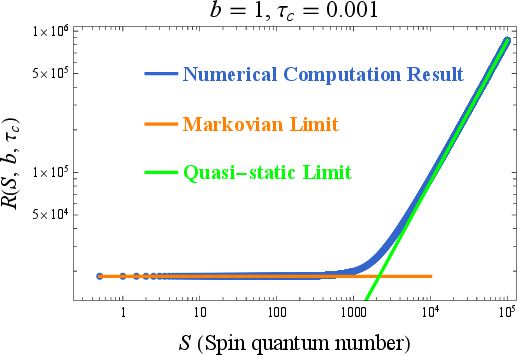}\caption{For fixed noise parameters
$(b,\tau_{c})$ with $b\tau_{c}=10^{-3}\ll1$ (i.e., Markovian noise for
spin-1/2), The behavior of $R$ as spin quantum number $S$ increases. The blue
curve represents the numerically computed values. The orange (green) curve
represents the scaling behavior of $R\propto S^{0}$ ($R\propto S$) in the
Markovian (quasi-static) regime.}%
\label{fig:sensitivity with s}%
\end{figure}

Using the QFI yield rate $R\sim\left(  2S\right)  ^{2}T_{2}$ or the estimation
precision per unit time $\delta\omega_{\min}\sqrt{T}=1/\sqrt{R} $ as the
figure of merit, we can now discuss the utility of the spin quantum number $S$
as a quantum resource (i.e., the scaling of the QFI yield rate $R$ with
respect to $S$) in different regimes (i.e., Markovian or non-Markovian) of the
noise. Recall that increasing $S$ amounts to increasing the signal $\omega$
and the noise $\tilde{\omega}(t)$ and $b$ by the same factor $S$.
Consequently, the scaling of $R\sim\left(  2S\right)  ^{2}T_{2}$ is determined
by the competition between the signal amplification factor $(2S)^{2}$ and the
decoherence time: $T_{2}\propto1/(Sb)$ (quasi-static noise) or $T_{2}%
\propto1/(Sb)^{2}$ (Markovian noise). In the absence of any noise, the spin
quantum number $S$ improves the estimation precision according to the
Heisenberg scaling $\delta\omega_{\min}\propto1/S$. In Markovian noises, the
spin quantum number $S$ cannot improve the estimation precision any more. In
quasi-static noises, the spin quantum number $S$ improves the estimation
precision according to the classical scaling $1/\sqrt{S}$. As shown in Fig.
\ref{fig:sensitivity with s}, for fixed noise parameters $(b,\tau_{c})$ with
$b\tau_{c}=10^{-3}\ll1$ (i.e., Markovian noise for spin-1/2), the scaling
behavior of $R$ gradually changes from $R\propto S^{0}$ in the Markovian
regime $S\ll1/(b\tau_{c})=10^{3}$ into $R\propto S\ $in the quasi-static
regime. When $b\tau_{c}\gg1$, the noise is always quasi-static and the QFI
yield rate $R\propto S$ for all $S$. 

\begin{figure}[ptb]
\centering
\includegraphics[width=\columnwidth]{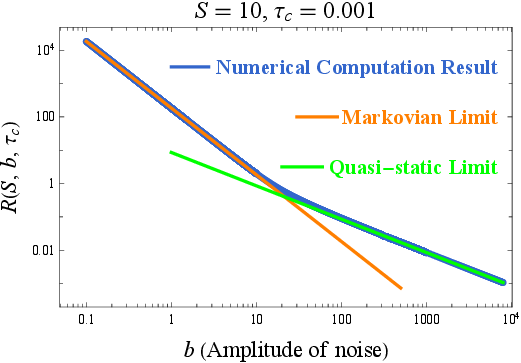}\caption{For fixed spin quantum
number $S$, the influence of the non-Markovianity increase (by increasing $b$)
of the noise on the QFI yield rate $R$. The blue curve represents the
numerically computed values. The orange (green) curve represents the scaling
behavior of $R\propto b^{-2}$ ($R\propto b^{-1}$) in the Markovian
(quasi-static) regime.}%
\label{fig:sensitivity with b}%
\end{figure}

\begin{figure}[ptb]
\centering
\includegraphics[width=\columnwidth]{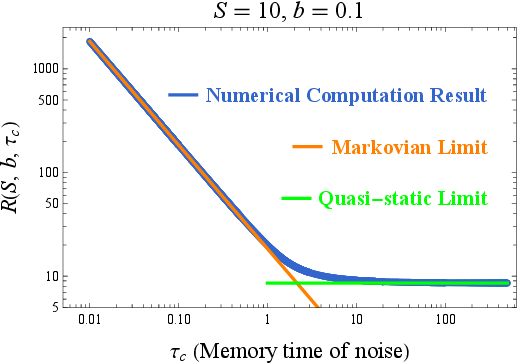}\caption{For fixed spin quantum
number $S$, the influence of the non-Markovianity increase (by increasing
$\tau_{c}$) of the noise on the QFI yield rate $R$. The blue curve represents
the numerically computed values. The orange (green) curve represents the
scaling behavior of $R\propto\tau_{c}^{-1}$ ($R\propto\tau_{c}^{0}$) in the
Markovian (quasi-static) regime.}%
\label{fig:sensitivity with tauc}%
\end{figure}

According to the discussions above, the spin quantum number $S$ is useful for
improving the estimation precision in non-Markovian noises, but is no longer
useful in Markovian noises. This may give us the mis-impression that
non-Markovian noises are \textquotedblleft better\textquotedblright\ than
Markovian ones for metrology. To clarify this point, we discuss how the noise
parameters influence the estimation precision. Since $R\sim S^{2}T_{2}$, the
influence of the noise on the estimation precision is completely contained in
$T_{2}$. According to Eq. (\ref{T2}), increasing the noise magnitude $b$
always decreases $T_{2}$, hence degrading the estimation precision. Increasing
the memory time $\tau_{c}$ first shortens $T_{2}$ (hence degrades the
estimation precision) in the Markovian regime $Sb\tau_{c}\ll1$ and then leaves
$T_{2}$ (hence the estimation precision) invariant when approaching the limit
of quasi-static noise $Sb\tau_{c}\gg1$. Therefore, when $S$ is fixed,
increasing the non-Markovianity (by increasing either $b$ or $\tau_{c}$) of
the noise will never improve the estimation precision. This is because in the
Markovian regime, the noise switches sign rapidly, so the accumulated random
phase grows up slowly, leading to slow spin decoherence. Increasing the noise
magnitude $b$ or its memory time $\tau_{c}$ (i.e., slows down the sign switch
of the noise) can only speed up the phase accumulation and hence the spin
decoherence. The influence of $b$ and $\tau_{c}$ on the QFI yield rate $R$ is
shown in Figs. \ref{fig:sensitivity with b} and
\ref{fig:sensitivity with tauc}.

6. \textit{Maintaining super-classical scaling in noisy environment. }
We might now wonder: can an estimation scheme achieve a precision dependence on quantum
resources $S$ that follows $1/S^{2}$? According to Eq.
(\ref{optimized QFI yield rate}), the QFI yield rate $R$ is proportional to
the product of $(2S)^{2}$ and the spin decoherence time $T_{2}$. Meanwhile,
according to Eq. (\ref{KAI_LIMIT}) and (\ref{spin decoherence time}), spin
decoherence time $T_{2}$ is related to $\chi\left(  \tau\right)  $. 

To achieve above goal, we can utilize Dynamical Decoupling (DD) \cite{PhysRev.80.580, yang2011preserving}, a potent technique for mitigating spin decoherence in a spin system. The key idea is to dynamically average out the coupling of the spin system to the environment by frequently flipping the spin system.

DD proves effective in mitigating decoherence when the DD-induced spin-flip occurs significantly faster than the memory time $\tau_c$ of the environmental noise. This ensures timely retrieval of lost coherence, preventing irreversible dissipation into the environment. However, in scenarios characterized by Markovian noise where $\tau \gg \tau_c$, the noise spectrum remains nearly constant across the entire filter bandwidth, thereby rendering DD ineffective.

Through DD, we can control $\chi\left(  \tau\right)$ such that it approximates:
\[
\chi\left(  \tau\right)  \sim\left\{
\begin{array}
[c]{ll}
\tau^{n} & \text{ \ }\left(  \tau\ll\tau_{c}\right)  ,\\
\tau & \text{ \ }\left(  \tau\gg\tau_{c}\right)  .
\end{array}
\right.
\]
Substituting the approximate expression for $\chi\left(  \tau\right)$ into Eq. (\ref{spin decoherence time}), we obtain the dependence of the QFI yield rate $R$ on the quantum resource $S$:
\[
R \sim (2S)^{2}T_{2} \sim\left\{
\begin{array}
[c]{ll}
S^{2-2/n} & \text{ \ }\left(  \tau\ll\tau_{c}\right)  ,\\
S^{0} & \text{ \ }\left(  \tau\gg\tau_{c}\right)  .
\end{array}
\right.
\]
So, by DD, we can achieve an estimation scheme where the precision dependence on quantum resources $S$ approaches $1/S^{2}$ in the quasi-static regime.

7. \textit{Numerical optimization for spin $S=1$: Maximizing final-state QFI. }
All the discussions above are based on adopting Eq. (\ref{GHZ}) as the initial
state of the spin-$S$ probe in the noisy case. This choice allows us to obtain
explicit analytical expressions for a clear physical picture on the utility of
the spin quantum number $S$ in different noises. However, athough the GHZ-like
state Eq. (\ref{GHZ}) is the optimal initial state in the absence of noises,
it may be no longer the case in the presence of noises. Here we provide a
prelimiary discussion about this point by considering $S=1$ and search
numerically for the optimal initial state that maximizes the final-state QFI.

For a spin-1, the pure initial state can be parameterized by four real
parameters $\Theta,\Phi,\lambda_{1},\lambda_{2}$:%
\[
\left\vert \psi\right\rangle =\cos\Theta\left\vert 1\right\rangle
+e^{i\lambda_{1}}\sin\Theta\cos\Phi\left\vert 0\right\rangle +e^{i\lambda_{2}%
}\sin\Theta\sin\Phi\left\vert -1\right\rangle .
\]
The noisy evolution drives the pure initial state into a mixed state
$\rho_{\omega}(\tau)$ after an interval $\tau$. Under the $S_{z}$ basis
$\left\vert 1\right\rangle $, $\left\vert 0\right\rangle $, $\left\vert
-1\right\rangle $, $\rho_{\omega}(\tau)$ is a $3\times3$ matrix:
\begin{widetext}
\begin{eqnarray}
\rho _{\omega }\left( \tau \right)  &=&  \notag \\
&&\left(
\begin{array}{ccc}
\cos ^{2}\Theta  & e^{-i\lambda _{1}}e^{-i\omega \tau }e^{-\chi \left(
\varsigma \right) }\cos \Theta \sin \Theta \cos \Phi  & e^{-i\lambda
_{2}}e^{-2i\omega \tau }e^{-4\chi \left( \varsigma \right) }\cos \Theta \sin
\Theta \sin \Phi  \\
e^{i\lambda _{1}}e^{i\omega \tau }e^{-\chi \left( \varsigma \right) }\cos
\Theta \sin \Theta \cos \Phi  & \sin ^{2}\Theta \cos ^{2}\Phi  & e^{i\left(
\lambda _{1}-\lambda _{2}\right) }e^{-i\omega \tau }e^{-\chi \left(
\varsigma \right) }\sin ^{2}\Theta \cos \Phi \sin \Phi  \\
e^{i\lambda _{2}}e^{2i\omega \tau }e^{-4\chi \left( \varsigma \right) }\cos
\Theta \sin \Theta \sin \Phi  & e^{-i\left( \lambda _{1}-\lambda _{2}\right)
}e^{i\omega \tau }e^{-\chi \left( \varsigma \right) }\sin ^{2}\Theta \cos
\Phi \sin \Phi  & \sin ^{2}\Theta \sin ^{2}\Phi   \notag
\end{array}\right)
\end{eqnarray}
\end{widetext}Using the method of Ref. \cite{vsafranek2018simple}, we obtain
the QFI of the final state as \begin{widetext}
\begin{eqnarray*}
\mathcal{F}(\omega) &\mathcal{=}&4\tau ^{2}e^{-8\chi }\sin ^{2}(\Theta )\sin
^{4}(\Phi ) \\
&&\times \frac{%
\begin{array}{c}
\cot ^{4}(\Theta )\left[ e^{12\chi }\cot ^{4}(\Phi )+2e^{6\chi }\left(
e^{4\chi }+e^{6\chi }+2\right) \cot ^{2}(\Phi )+4\left( e^{2\chi }+e^{4\chi
}+e^{6\chi }+1\right) \right]  \\
+2e^{6\chi }\cot ^{2}(\Theta )\cos ^{2}(\Phi )\left[ e^{2\chi }\left(
2e^{2\chi }+e^{4\chi }-2\right) \cot ^{2}(\Phi )+e^{4\chi }+e^{6\chi }+2%
\right] +e^{12\chi }\cos ^{4}(\Phi )%
\end{array}%
}{%
\begin{array}{c}
\left( e^{2\chi }+e^{4\chi }+e^{6\chi }+1\right) \cot ^{2}(\Theta )\sin
^{2}(\Phi )\left[ \cot ^{2}(\Theta )+\sin ^{2}(\Phi )\right] +e^{6\chi }\cos
^{4}(\Phi )\left[ \cot ^{2}(\Theta )+\sin ^{2}(\Phi )\right]  \\
+e^{2\chi }\cos ^{2}(\Phi )\left[ 2\left( e^{2\chi }+e^{4\chi }+1\right)
\cot ^{2}(\Theta )\sin ^{2}(\Phi )+e^{4\chi }\cot ^{4}(\Theta )+e^{4\chi
}\sin ^{4}(\Phi )\right]
\end{array}}
\end{eqnarray*}
\end{widetext}The final-state QFI depends on the spin quantum number $S$, the
noise parameters $b,\tau_{c}$, and two initial state parameters $\Theta,\Phi$,
but is independent of $\lambda_{1}$ and $\lambda_{2}$, because the relative
phase between $S_{z}$ eigenstates in the initial state can be generated by
$e^{-i\phi_{2}S_{z}^{2}}e^{-i\phi_{1}S_{z}}$ with suitable $\phi_{1}$ and
$\phi_{2}$ and these operations commute with the noisy evolution. Next we
optimize $\tau\ $to obtain the QFI yield rate $R\equiv\max_{\tau}%
\mathcal{F}\left(  \tau\right)  /\tau$, which still depends on $\Theta
,\Phi,S,b,\tau_{c}$. Finally, we optimize the initial-state parameters
$\Theta,\Phi$ to obtain $R_{\max}$ as a function of $S,b,\tau_{c}$.

\begin{figure}[ptb]
\centering
\includegraphics[width=\columnwidth]{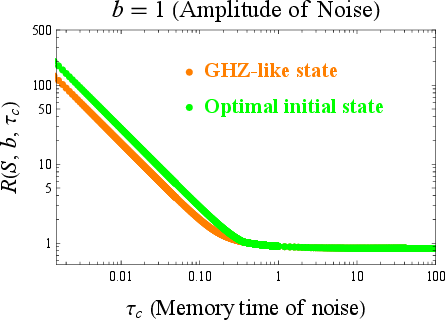}
\caption{The impact of optimizing the initial state on $R\left( S,b,\tau_{c}\right)  $ when $b=1$}
\label{fig:The impact of optimizing the initial state when b=1}
\end{figure}

\begin{figure}[ptb]
\centering
\includegraphics[width=\columnwidth]{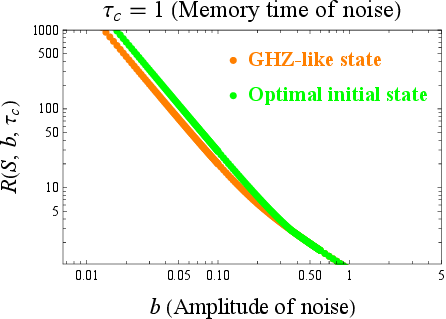}
\caption{The impact of optimizing the initial state on $R\left( S,b,\tau_{c}\right)  $ when $\tau_{c}=1$}
\label{fig:The impact of optimizing the initial state when tauc=1}
\end{figure}

\begin{figure}[hptb]
\centering
\includegraphics[width=\columnwidth]{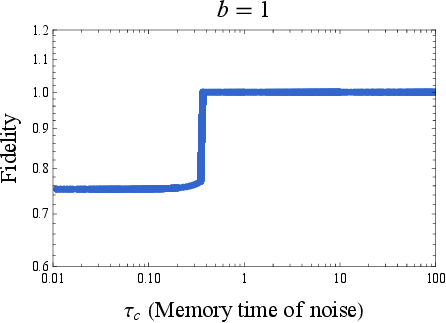}\caption{Fidelity between
the GHZ-like state and the optimal initial state as a function of $\tau_{c}$
when $b=1$}%
\label{fig:fidelity}%
\end{figure}

According to the numerical results shown in Figs.
\ref{fig:The impact of optimizing the initial state when b=1} and
\ref{fig:The impact of optimizing the initial state when tauc=1}, for
quasi-static noises $Sb\tau_{c}\gg1$, the GHZ-like state Eq. (\ref{GHZ}) is
almost the optimal initial state. The similarity between the GHZ-like state
and the optimal initial state can be quantified by computing their fidelity,
which represents the absolute value of the inner product between the two
states. It's noteworthy that while the QFI remains independent of the relative
phase in the initial state, fidelity depends on these parameters in both
states. Therefore, when calculating the fidelity between the GHZ-like state
and the optimal initial state, it becomes a function of phase parameters. To
gauge the highest level of similarity between the GHZ-like state and the
optimal initial state, we maximize the fidelity with respect to phase
parameters, i.e., setting phase parameters to zero. The specific computation
of fidelity between the GHZ-like state and the optimal initial state is
depicted in Fig. \ref{fig:fidelity}. So optimizing the initial state provides
almost no improvement of $R$ compared with the GHZ-like initial state. A very
rough understanding is as follows. In quasi-static noises, the initial state
$(|S\rangle+|-S\rangle)/\sqrt{2}$ composed of the highest and lowest
eigenstates of $S_{z}$ is better than the initial state $(|m\rangle
+|-m\rangle)/\sqrt{2}$ ($|m|<S$) composed of intermediate eigenstates of
$S_{z}$. This implies that introducing intermediate eigenstates of $S_{z}$
into the GHZ-like state may not improve the estimation precision. In the
opposite limit of Markovian noises $Sb\tau_{c}\ll1$, the optimal initial state
differs appreciably from the GHZ-like state and improves the QFI yield rate by
a constant factor.

8. \textit{Conclusion}
Through the computational analysis in this study, we have found that in the
absence of noise, the standard deviation $\delta\omega_{\min}$ of the optimal
estimation scheme is related to the spin quantum number $S$, proportional to
$1/S$. When noise is present, using the QFI yield rate $R\sim\left(
2S\right)  ^{2}T_{2}$, we discussed the utility of the spin quantum number $S$
as a quantum resource in different noise regimes (i.e., Markovian or
non-Markovian). The scaling of $R\sim\left(  2S\right)  ^{2}T_{2} $ is
determined by the competition between the signal amplification factor $\left(
2S\right)  ^{2}$ and the decoherence time: $T_{2}\propto1/(Sb)$ (quasi-static
noise) or $T_{2}\propto1/(Sb)^{2}$ (Markovian noise). In Markovian noises, the
spin quantum number $S$ cannot further improve the estimation precision.
However, in quasi-static noises, the spin quantum number $S$ enhances the
estimation precision according to the classical scaling $1/\sqrt{S}$. In other
words, the spin quantum number $S$ is beneficial for improving estimation
precision in non-Markovian noises but loses its effectiveness in Markovian noises.

\bigskip

\begin{acknowledgments}
This work was supported by the National Natural Science Foundation of China
(NSFC) Grant No. 12274019 and the NSAF grant in NSFC with grant No. U2230402.
We acknowledge the computational support from the Beijing Computational
Science Research Center (CSRC).
\end{acknowledgments}

\bibliography{references}

\end{document}